\documentclass[letter, amsfonts, amssymb, amsmath, reprint, showkeys, nofootinbib, twoside, superscriptaddress]{revtex4-1}

\usepackage{color}
\usepackage{amsmath}
\usepackage{pifont}   
\usepackage{verbatim}       
\usepackage{acronym} 
    
\usepackage{amsfonts}
\usepackage{amssymb}
\usepackage{mathrsfs} 
\usepackage{graphicx}
\usepackage{multirow} 
\usepackage{slashed} 
\usepackage{array}

\usepackage{graphicx}
\usepackage{epsf}
\usepackage{bm}
\usepackage{booktabs}
\usepackage{array}
\usepackage{caption}
\usepackage{subcaption}
\usepackage[utf8]{inputenc}

\allowdisplaybreaks

\usepackage[pdftex, pdftitle={Article}, pdfauthor={Author}]{hyperref} 

\begin{document}

\title{Is the decay of the Higgs boson to a photon and a dark photon currently observable at the LHC?}

\author{Hugues Beauchesne}
  \email[Email address: ]{beauchesneh@phys.ncts.ntu.edu.tw}
  \affiliation{Physics Division, National Center for Theoretical Sciences, Taipei 10617, Taiwan}

\author{Cheng-Wei Chiang}
  \email[Email address: ]{chengwei@phys.ntu.edu.tw}
  \affiliation{Department of Physics and Center for Theoretical Physics, National Taiwan University, Taipei 10617, Taiwan}
  \affiliation{Physics Division, National Center for Theoretical Sciences, Taipei 10617, Taiwan}

\date{\today}

\begin{abstract}
Many attempts have been made to observe the decay of the Higgs boson to a photon and an invisible massless dark photon. For this decay to be potentially observable at the LHC, new mediators that communicate between the Standard Model and the dark photon must exist. In this Letter, we study bounds on such mediators coming from the Higgs signal strengths, oblique parameters, electric dipole moment of the electron and unitarity. We find that the branching ratio of the Higgs boson to a photon and a dark photon is constrained to be far smaller than the sensitivity of current collider searches, thus calling for a reconsideration of current experimental efforts.
\end{abstract}

\maketitle

{\it Introduction -- }
A dark photon is a new Abelian gauge boson that can mix with the photon \cite{Holdom:1985ag} and has been the subject of extensive theoretical and experimental studies. One potential discovery channel that has been the subject of much scrutiny is the decay of the Higgs boson to a photon and a massless invisible dark photon, $h \to A A'$. Searching for this decay was first motivated by a series of phenomenological papers \cite{Gabrielli:2014oya, Biswas:2015sha, Biswas:2016jsh, Biswas:2017lyg, Biswas:2017anm}, which claimed that the branching ratio of the Higgs to a photon and an invisible particle could be as high as 5\% and still be compatible with experimental constraints at the time. This led to a series of experimental searches at the LHC \cite{CMS:2019ajt, CMS:2020krr, ATLAS:2021pdg}. At the time of writing this Letter, the strongest limit on this branching ratio comes from Ref.~\cite{ATLAS:2021pdg}, which obtains a bound of 1.8\% at 95\% confidence limit (CL).

The decay of the Higgs boson to a photon and a dark photon could in principle simply be the result of tree-level interactions of the Standard Model (SM) particles with the dark photon. The problem with this scenario however is that the interactions between the SM particles and a new light gauge boson are constrained to be very small. If the decay of $h \to A A'$ is to be realistically observable at the LHC, new particles that mediate interactions between the SM particles and the dark photon must therefore exist.

In this Letter, we investigate experimental and theoretical constraints on mediators that allow the $h \to A A'$ process. The constraints considered are the Higgs signal strengths, oblique parameters, electric dipole moment (EDM) of the electron, and unitarity. Most importantly, we demonstrate that these constraints restrict the Higgs branching ratio to a photon and a dark photon, $\text{BR}(h \to A A')$, to be far smaller than the sensitivity of current collider searches barring extremely fine-tuning or contrived model building.

Unfortunately, it is not possible to obtain a bound on $\text{BR}(h \to A A')$ that is truly model independent. Certain observables like the Higgs signal strengths are simply too complicated and can be affected in too many ways. As such, we will study a large and representative set of benchmark models. Although they are only benchmarks, they will illustrate clearly why obtaining a $\text{BR}(h \to A A')$ potentially observable at the LHC would require significant fine-tuning.

In the models we consider, $\text{BR}(h \to A A')$ is always constrained to be below $\sim 0.4\%$. Furthermore, this upper limit could only be realized in the presence of light charged mediators that somehow would have avoided experimental constraints. The improvement over previous limits comes from a more accurate analysis of the Higgs signal strengths, the inclusion of new constraints (oblique parameters, EDM of the electron and unitarity) and the inclusion of more recent experimental data.


{\it Benchmark models -- }
We begin by presenting the models considered in this Letter. Consider a new $U(1)'$ gauge group whose gauge boson is $A'$ and under which SM particles are neutral. Assume a set of mediator fields that are charged under both SM gauge groups and $U(1)'$. We then confine ourselves to the models satisfying the following requirements:
\begin{enumerate}
  \item Have a renormalizable Lagrangian that preserves all gauge symmetries;
  \item Lead to the $h \to AA'$ decay at one loop;
  \item Contain no mediators charged under QCD;
  \item Contain only mediators that are complex scalars or vector-like fermions;
  \item Contain no more than two new fields; and
  \item Contain no mediators that mix with SM fields or have a non-zero expectation value.
\end{enumerate}
Requirement 2 is made since a decay at multiple loops would not lead to an observable $\text{BR}(h \to AA')$. Requirement 3 is imposed because a coloured mediator would lead to a large modification to the gluon-fusion cross section. Requirements 4 to 6 are imposed to keep the number of possible models to a manageable level.

In practice, these requirements simply mean that the Lagrangian must include a term involving both the Higgs doublet $H$ and the mediators. Conveniently, the vertices that satisfy these requirements fall into a finite number of categories. They are:

1) Fermion case:
\begin{equation}\label{eq:CaseILagrangian}
  \mathcal{L}^F =-\hat{d}^{pn}_{abc} \bar{\psi}_1^a (A_L P_L + A_R P_R) \psi_2^b H^c + \text{h.c.},
\end{equation}
where the $\psi_i$ are fermions with gauge numbers
\begin{equation}\label{eq:CaseFGaugeNumbers}
  \psi_1:\; (\mathbf{1}, \mathbf{p}, Y^p, Q'), \quad 
  \psi_2:\; (\mathbf{1}, \mathbf{n}, Y^n, Q'),
\end{equation}
with $Y^p = Y^n + 1/2$, $p = n \pm 1$ and the gauge numbers written in the form $(SU(3), SU(2)_L, U(1)_Y, U(1)')$. The quantity $\hat{d}^{pn}_{abc} $ is an $SU(2)_L$ tensor and can be expressed in terms of Clebsch-Gordan (CG) coefficients.

2) Scalar case I:
\begin{equation}\label{eq:CaseSILagrangian}
  \mathcal{L}^{S_I} = -\mu \hat{d}_{abc}^{pn} \phi_1^{a\dagger}\phi_2^b H^c + \text{h.c.},
\end{equation}
where the $\phi_i$ are scalars with gauge numbers
\begin{equation}\label{eq:CaseSIGaugeNumbers}
  \phi_1:\; (\mathbf{1}, \mathbf{p}, Y^p, Q'), \quad
  \phi_2:\; (\mathbf{1}, \mathbf{n}, Y^n, Q'),
\end{equation}
with $Y^p = Y^n + 1/2$.

3) Scalar case II:
\begin{equation}\label{eq:CaseSIILagrangian}
  \mathcal{L}^{S_{II}} = -\sum_r \lambda^r \hat{d}_{abcd}^{nr} H^{a\dagger} H^b \phi^{c\dagger}\phi^d,
\end{equation}
where $\phi$ is a scalar with gauge numbers
\begin{equation}\label{eq:CaseSIIGaugeNumbers}
  \phi:\; (\mathbf{1}, \mathbf{n}, Y, Q'),
\end{equation}
$r$ refers to different ways of contracting $SU(2)_L$ indices, and $\hat{d}_{abcd}^{nr}$ are $SU(2)_L$ tensors which can be expressed as a combination of CG coefficients. If $n \neq 1$, there are two possible contractions each with its own coefficient $\lambda^r$.

4) Scalar case III:
\begin{equation}\label{eq:CaseSIIILagrangian}
  \mathcal{L}^{S_{III}} = -\sum_r \lambda^r \hat{d}_{abcd}^{pnr} H^{a\dagger} H^b \phi^{c\dagger}_1\phi^d_2 + \text{h.c.},
\end{equation}
where the $\phi_i$ are scalars with gauge numbers
\begin{equation}\label{eq:CaseSIIIGaugeNumbers}
  \phi_1:\; (\mathbf{1}, \mathbf{p}, Y^p, Q'), \quad
  \phi_2:\; (\mathbf{1}, \mathbf{n}, Y^n, Q'),
\end{equation}
where $Y^p = Y^n$ and $p \in \{n - 2, n, n + 2\}$. If $p = n \neq 1$, there are two ways to contract the $SU(2)_L$ indices and one way otherwise.

5) Scalar case IV:
\begin{equation}\label{eq:CaseSIVLagrangian}
  \mathcal{L}^{S_{IV}} = -\lambda \hat{d}_{abcd}^{pn} H^a H^b \phi^{c\dagger}_1\phi^d_2 + \text{h.c.},
\end{equation}
where the $\phi_i$ are scalars with gauge numbers
\begin{equation}\label{eq:CaseSIVGaugeNumbers}
  \phi_1:\; (\mathbf{1}, \mathbf{p}, Y^p, Q'), \quad
  \phi_2:\; (\mathbf{1}, \mathbf{n}, Y^n, Q'),
\end{equation}
where $Y^p = Y^n + 1$ , $p \in \{n - 2, n, n + 2\}$ and $p$ and $n$ are assumed not to both be 1. There is only a single way to contract the $SU(2)_L$ indices.

The different models could of course be combined, but we will assume for manageability sake that only one type of vertex is present. Once the Higgs field $H$ obtains an expectation value, the mediators will usually mix. The only exception to this is scalar case II. The interactions with the Higgs boson and $Z$ boson are usually non-trivial.

{\it Constraints -- }
We now consider the constraints to be imposed on the different models introduced above.

{\it A. Higgs signal strengths: }
Gauge invariance forces the amplitude of the decay $h \to AA$ to take the form
\begin{equation}\label{eq:HiggsDecayI}
  \begin{aligned}
    M^{h\to AA} = & S^{h\to AA} \left(p_1\cdot p_2 g_{\mu\nu} - p_{1\mu} p_{2\nu}\right)\epsilon^\nu_{p_1}\epsilon^\mu_{p_2}\\
                & + i\tilde{S}^{h\to AA}\epsilon_{\mu\nu\alpha\beta}p_1^\alpha p_2^\beta\epsilon^\nu_{p_1}\epsilon^\mu_{p_2}.\\
  \end{aligned}
\end{equation}
The amplitudes for $AA'$ and $A'A'$ have similar forms. At one loop, the $S$ coefficients are
\begin{equation}\label{eq:HiggsDecayII}
  \begin{aligned}
    S^{h\to AA  } &= e^2    \sum_a Q_a^2  C_a + S^{h\to AA}_\text{SM},\\
    S^{h\to AA' } &= e e'   \sum_a Q_a Q' C_a,                        \\
    S^{h\to A'A'} &= {e'}^2 \sum_a {Q'}^2            C_a,
  \end{aligned}
\end{equation}
where $e'$ is the gauge coupling constant of $U(1)'$, the sum is over the different mediators, $C_a$ is a common factor, and $S^{h\to AA}_\text{SM}$ is the SM contribution. Coefficients of the type $\tilde{S}$ take analogous forms. It is clear from Eq.~\eqref{eq:HiggsDecayII} that a large $\text{BR}(h \to A A')$ will generally lead to either a large $\text{BR}(h \to A' A')$, a large modification of $\text{BR}(h \to A A)$, or both. The Higgs signal strengths thus play a crucial role in constraining $\text{BR}(h \to A A')$.

The Higgs signal strength constraints are applied using the $\kappa$ formalism \cite{Heinemeyer:2013tqa}. The only two affected at leading order are those associated to $AA$ and $AZ$. The decays to $A'A'$, $A A'$ and $Z A'$ are taken into account by properly rescaling the signal strengths. This global reduction of the Higgs signal strengths renders the searches for the Higgs decaying to invisible particle superfluous. A global fit is then performed by using the most up-to-date measurements of the Higgs signal strengths of Refs.~\cite{CMS-PAS-HIG-19-005} and \cite{ATLAS-CONF-2021-053} by CMS and ATLAS, respectively. These references conveniently provide the measurements, uncertainties and correlations of the Higgs signal strengths. We impose the bounds at $95\%$ CL.\footnote{It could technically be possible for the mediator contributions to $S^{h\to AA  }$ to be about $-2S^{h\to AA}_\text{SM}$.This would avoid most of the signal strength constraints. It would however require exotic circumstances and a large amount of fine tuning. We will ignore such cases.}

{\it B. Electron EDM: }
There is a potential loophole in the constraints of the Higgs signal strengths. The decay width of the Higgs to two photons is proportional to $|S^{h\to AA}|^2 + |\tilde{S}^{h\to AA}|^2$. In general, there will be interference between the contributions from the SM particles and the new mediators. These terms are generally responsible for most of the modification to $\text{BR}(h \to A A)$. The interference terms could potentially be avoided in two ways. First, the SM contribution to $S^{h\to AA}$ is almost purely real. Interference terms could then be avoided by having a purely imaginary mediator contribution to $S^{h\to AA}$. However, it is easy to verify that $C_a$ is forced to be purely real for charged fermion and scalar mediators. The only way it could be complex would be for the mediators to be lighter than $m_h/2$; but this is in blatant violation of LEP bounds \cite{LEP1, LEP2}. Second, the SM contribution to $\tilde{S}^{h\to AA}$ is essentially 0. Interference terms could have been avoided by having the mediators exclusively contribute to $\tilde{S}^{h\to AA}$. For the scalars, this is not possible. However, this can be done for the fermions and the limits from the Higgs signal strengths could be greatly weakened. Thankfully, a side effect of this scenario would be a large contribution to the EDM of the electron, which is strongly constrained.

The contributions to the electron EDM come from Barr-Zee diagrams involving a combination of the photon, $Z$ boson, $W$ boson and the Higgs boson. The contributions of the diagrams involving a Higgs boson and either a photon or $Z$ are taken from Ref.~\cite{Nakai:2016atk}. The contribution of diagrams involving two $W$ bosons can be computed by adapting the results of Ref.~\cite{Chang:2005ac}. The limit on the electron EDM of Ref.~\cite{ACME:2018yjb} by the ACME collaboration is used and corresponds to $90\%$ CL.

Once the electron EDM constraints are implemented, $\tilde{S}^{h\to AA}$ will be forced to be essentially zero for a $\text{BR}(h \to A A')$ close to its limit. There is therefore no point in considering other $CP$-violating observables.

{\it C. Oblique parameters: }
The operators presented in the benchmark models generally lead to different masses for particles that are part of the same representation of the electroweak groups. This leads to modifications of the oblique parameters $S$ and $T$ \cite{Peskin:1990zt}. For the fermion case, we use the general results of Ref.~\cite{Anastasiou:2009rv, Lavoura:1992np, Chen:2003fm, Carena:2007ua, Chen:2017hak, Cheung:2020vqm}. For the scalar case, we compute them ourselves. The limits of Ref.~\cite{Zyla:2020zbs} are used and correspond to $95\%$ CL.

{\it D. Unitarity: }
Unitarity imposes bounds on the coefficients in Eqs.~\eqref{eq:CaseILagrangian}, \eqref{eq:CaseSIILagrangian}, \eqref{eq:CaseSIIILagrangian}, and \eqref{eq:CaseSIVLagrangian}. For the fermion case, unitarity of the scattering $\bar{\psi}_1^a\psi_2^b \to \bar{\psi}_1^c\psi_2^d$ requires
\begin{equation}\label{eq:Case1a0Limit2}
  |A_R|^2 + |A_L|^2 < \frac{32\pi}{p}.
\end{equation}
For the scalar cases, unitarity of the scattering of two mediators to two Higgs bosons requires
  \begin{align}
    \text{case II:} \quad &\frac{1}{16\sqrt{2}\pi}         \left[\sum_{i,j} \left|\sum_r \lambda^r \hat{d}^{nr}_{22ij}  \right|^2 \right]^{\frac{1}{2}} &< \frac{1}{2},\nonumber\\
    \text{case III:}\quad &\frac{1}{16\sqrt{2}\pi}         \left[\sum_{i,j} \left|\sum_r \lambda^r \hat{d}^{pnr}_{22ij} \right|^2 \right]^{\frac{1}{2}} &< \frac{1}{2},\\
    \text{case IV:} \quad &\frac{|\lambda|}{16\sqrt{2}\pi} \left[\sum_{i,j} \left|                 \hat{d}^{pn}_{22ij}  \right|^2 \right]^{\frac{1}{2}} &< \frac{1}{2}.\nonumber
  \end{align}
For the scalar case IV, this can be simplified to
\begin{equation}\label{eq:CaseIIUnitarityII}
  |\lambda| < 8 \pi \sqrt{\frac{6}{p}}.
\end{equation}
No unitarity bound generally applies on $\mu$ for scalar case I. For the scalar cases, we also apply the unitarity bounds on $Q'e'$ by adapting the results of Ref.~\cite{Hally:2012pu}. This gives
\begin{equation}\label{eq:Unitarityepscalar}
    |Q'e'| < \frac{\sqrt{4\pi}}{q^{1/4}},
\end{equation}
where $q= n + p$ for cases I, III, IV and $q = n$ for case II.

\begin{figure*}[t!]
\begin{minipage}[b]{.97\textwidth}
\begin{center}
 \begin{subfigure}{0.49\textwidth}
    \centering
    \includegraphics[width=1\textwidth]{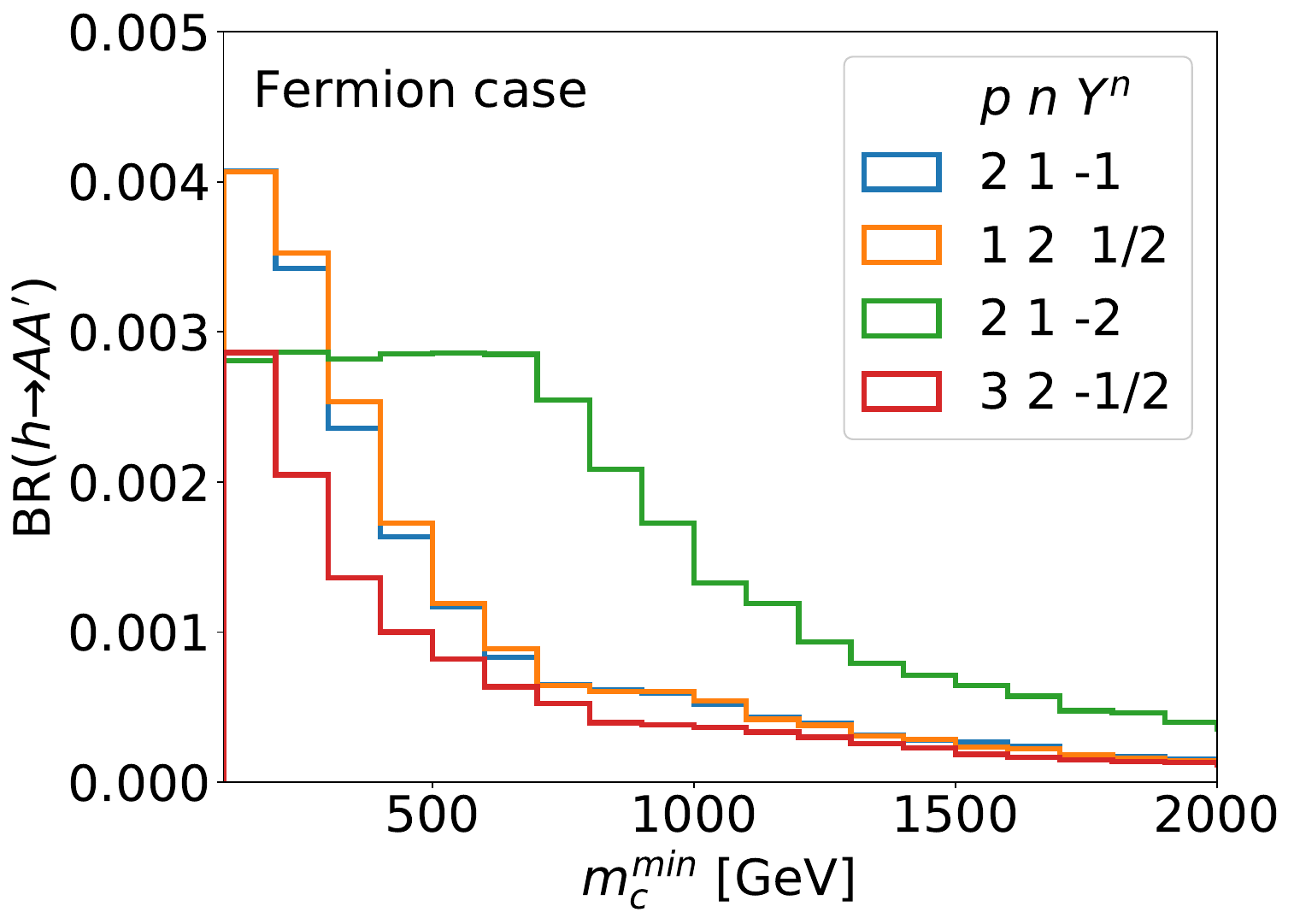}
    \label{fig:Fermion}
 \end{subfigure}
 \begin{subfigure}{0.49\textwidth}
    \centering
    \includegraphics[width=1\textwidth]{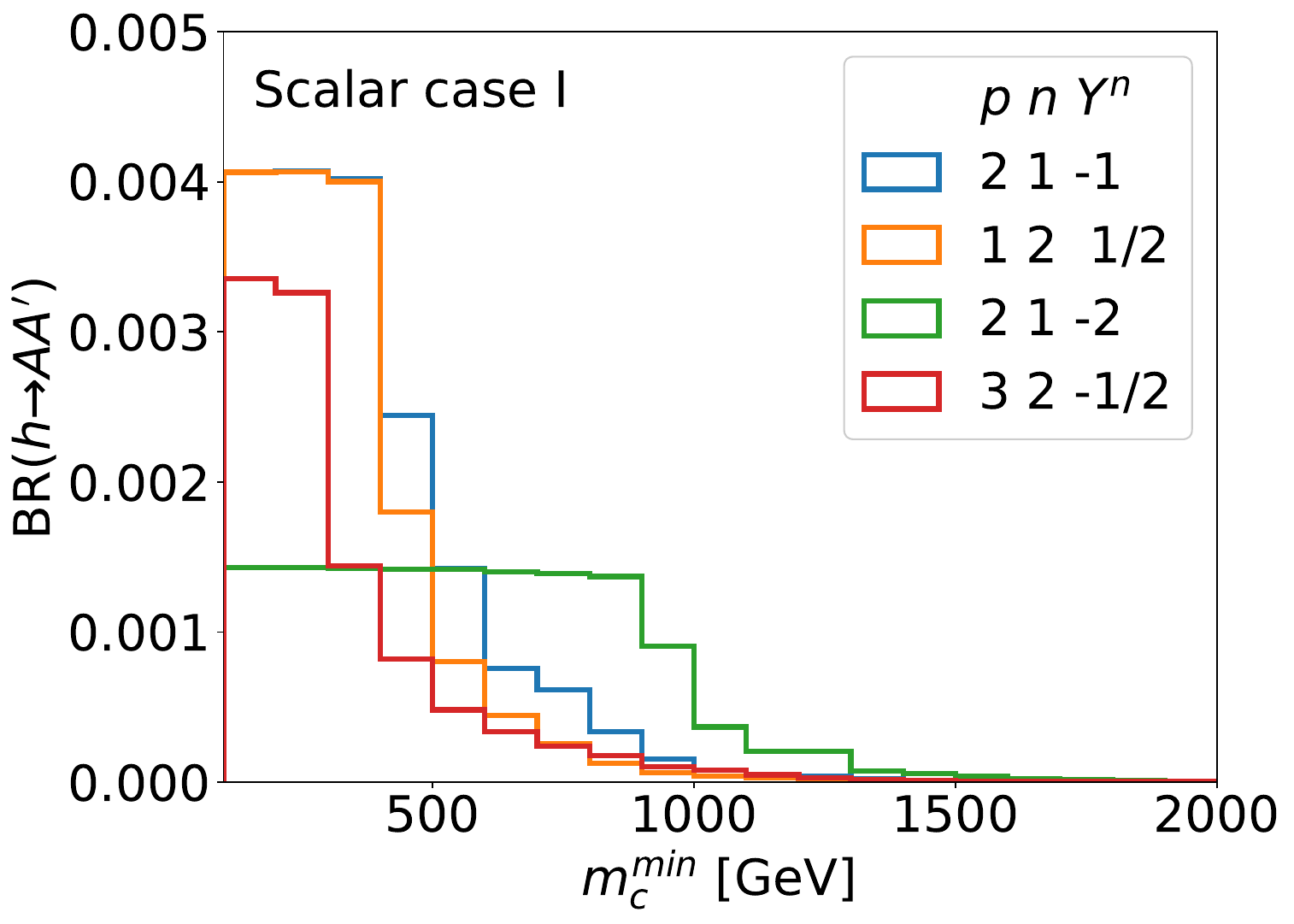}
    \label{fig:Scalar_I}
 \end{subfigure}
 \begin{subfigure}{0.49\textwidth}
    \centering
    \includegraphics[width=1\textwidth]{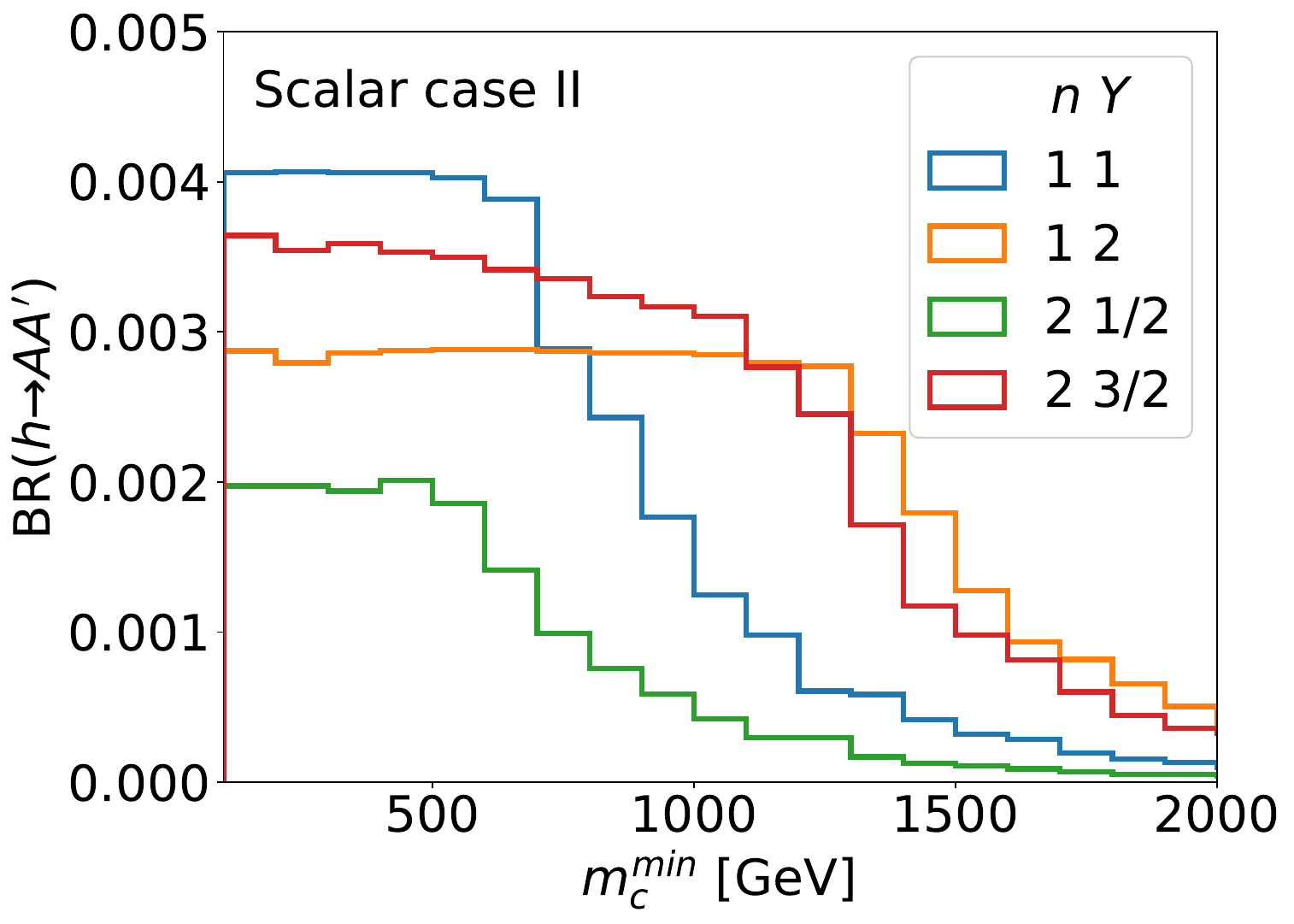}
    \label{fig:Scalar_II}
 \end{subfigure}
 \begin{subfigure}{0.49\textwidth}
    \centering
    \includegraphics[width=1\textwidth]{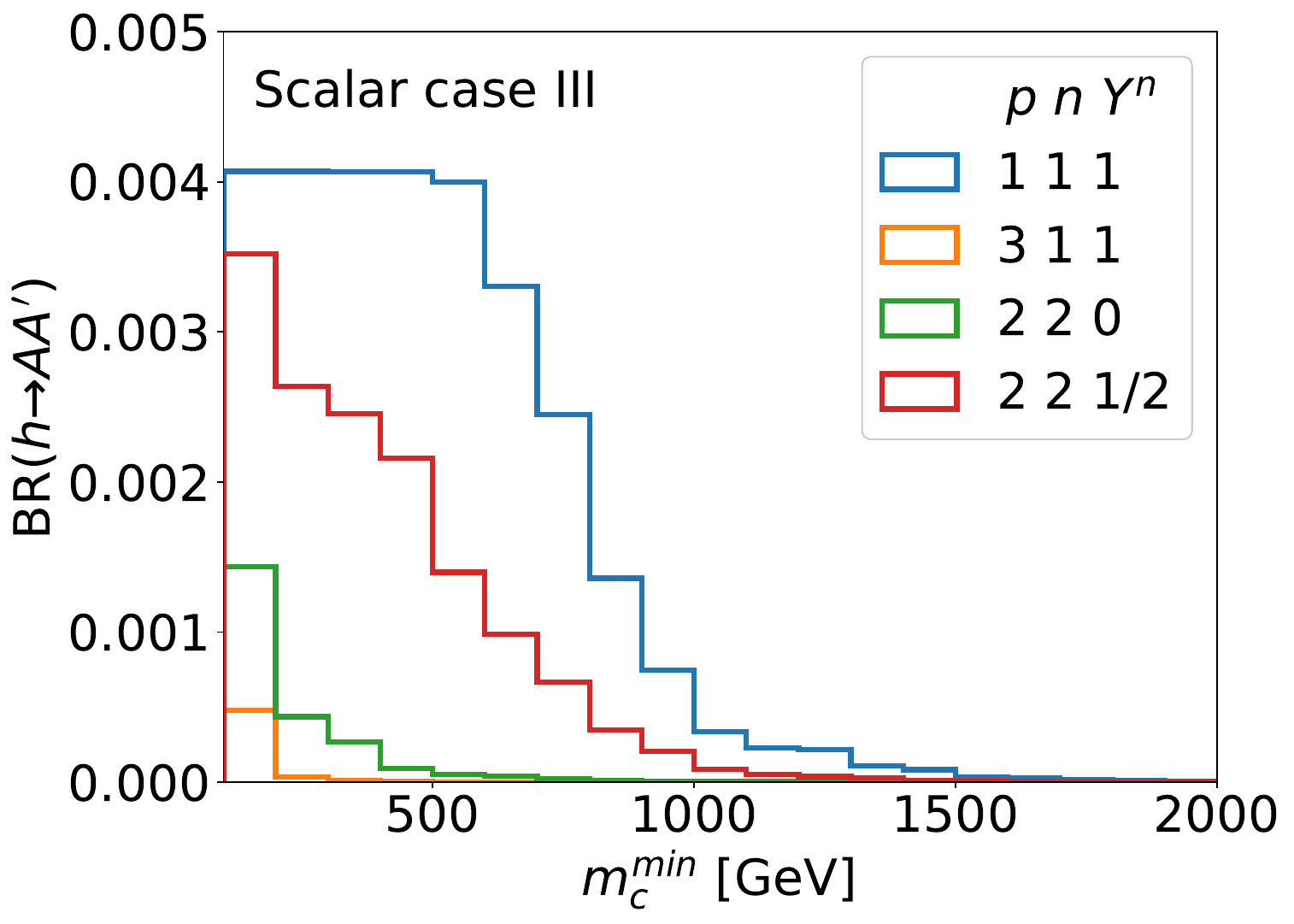}
    \label{fig:Scalar_III}
 \end{subfigure}
\caption{Maximum allowed $\text{BR}(h \to A A')$ for different examples of mediator models. The plots do not go below 100~GeV, as LEP bounds would prohibit charged particles of such masses \cite{LEP1, LEP2}.}
\label{fig:ThePlot}
\end{center}
\end{minipage}
\end{figure*}

{\it Results -- }
We now present the limits on $\text{BR}(h \to A A')$ for the benchmark models. In all cases considered, the parameter space is scanned by using a Markov Chain with the Metropolis-Hasting algorithm. To maximize the number of points near the limits, a prior proportional to $\text{BR}(h \to A A')^2$ is used. As the results only depend on $Q'$ and $e'$ via the product $Q'e'$, the limits are independent of the choice of $Q'$ and we impose $|Q'e'| < \sqrt{4\pi}$. For each model, the results are finely binned and the largest $\text{BR}(h \to A A')$ is selected in each bin. The results are shown in Fig.~\ref{fig:ThePlot}. The mass $m_c^{\text{min}}$ denotes the mass of the lightest charged mediator.

As can be seen, $\text{BR}(h \to A A')$ never exceeds $\sim 0.4\%$ for any of the models considered. A limit of this order of magnitude is easy to predict. Consider Eq.~\eqref{eq:HiggsDecayII} and assume that $C_a$ is real and $\tilde{S}^{h\to AA} = 0$ as argued above. Further, assume that the contributions to the Higgs decay to $AA$, $AA'$ and $A'A'$ are all dominated by a single mediator. It is then easy to prove that
\begin{equation}\label{eq:CaseIAmplitudeRelation}
  \begin{aligned}
    & \text{BR}(h \to AA') \approx \\
    & \hspace{0.5 cm} \sqrt{\text{BR}(h\to A'A') \text{BR}(h\to AA)} \left|\frac{\Delta \text{BR}(h \to AA)}{\text{BR}(h\to AA)}\right|,
  \end{aligned}
\end{equation}
where $\Delta \text{BR}(h \to AA)$ is the deviation of $\text{BR}(h \to AA)$ from its SM value. The Higgs decay to invisible particle $\text{BR}(h\to A'A')$ can be at most $\mathcal{O}(10\%)$. The branching ratio to two photons $\text{BR}(h\to AA)$ is about $0.23\%$ and can at most deviate from this value by $\mathcal{O}(25\%)$. One then expects $\text{BR}(h \to AA')$ to be at most $\mathcal{O}(0.4\%)$. Of course, there are in general multiple particles in the loop and there can be small deviations from this number. However, obtaining a $\text{BR}(h\to AA')$ much higher than $0.4\%$ would require some precise cancellations to take place in $h \to AA$ and $h \to A'A'$ decays. Plus, interactions with the Higgs will often considerably split the masses of the mediators and force a single mediator to dominate.

Many models show a plateau of maximum $\text{BR}(h\to AA')$ followed by a decreasing limit. The plateau is due to the Higgs signal strengths. The threshold at which this plateau ends corresponds to where the constraints from the oblique parameters or unitarity become stronger than the Higgs signal strengths constraints.

The figures clearly indicate that the limits depend on the mass of the lightest charged mediator. In principle, there should be a lower limit on the mass of such particles from collider searches. This would tighten the constraints on $\text{BR}(h\to AA')$. In practice, the mediators could have complicated decays that are not searched for and a lower limit cannot be technically applied besides the LEP bound. However, it is very improbable that a charged particle of less than a few hundred GeV would not have been found at the LHC by now.

Some mediators are more strictly constrained and cannot even reach a $\text{BR}(h\to AA')$ of $0.4\%$. This is namely the case for the mediators of scalar case IV. These lead to negative contributions to the $T$ oblique parameter and can only result in a very small $\text{BR}(h\to AA')$, which is why we do not include any plot for this case. Obtaining a large $\text{BR}(h\to AA')$ typically requires $|Q'e'|$ to be much larger than $|Q e|$, where $Q$ is the electric charge of a mediator. This is problematic for mediators with large electric charge, as $|Q'e'|$ is bounded from above.

Scalar case II and scalar case III for $p=n$ can avoid contributions to the oblique parameters by an appropriate choice of coefficients. This however results in degenerate mediator masses. For scalar case II in particular, it can be shown that degenerate masses lead to
\begin{equation}\label{eq:CaseIIIPlateaus}
  \begin{aligned}
    & \text{BR}(h \to AA') \approx \frac{1}{1 + \frac{n^2 - 1}{12 Y^2}}\\
    & \hspace{0.5cm} \times\sqrt{\text{BR}(h\to A'A') \text{BR}(h\to AA)} \left|\frac{\Delta \text{BR}(h \to AA)}{\text{BR}(h\to AA)}\right|.
  \end{aligned}
\end{equation}

{\it Conclusion -- }
In this Letter, we studied constraints on mediators that allowed the Higgs boson to decay to a photon and an invisible massless dark photon. To do so, we considered a large and representative set of benchmarks models. We found that constraints from the Higgs signal strengths, EDM of the electron, oblique parameters and unitarity forced $\text{BR}(h\to AA')$ to be at most $\sim 0.4\%$ for these models. This is far below the current collider bound of $1.8\%$. Furthermore, this would require the presence of light charged particles which would somehow have avoided detection.

In addition, obtaining a sizable $\text{BR}(h\to AA')$ requires the mediators to satisfy certain requirements that may not be very aesthetically pleasing. Namely, they require very large couplings with the Higgs and a large dark electric charge. In the case of Yukawa couplings, they are often required to be of order a few. Worse, the models always lead to a Landau pole at low energy, sometimes as low as the TeV scale.

There could in principle be ways to obtain a larger $\text{BR}(h\to AA')$. This could be done for example by including different combinations of the models of this Letter and carefully adjusting them to cancel the contributions to the Higgs decay to $AA$ or $A'A'$. However, obtaining a $\text{BR}(h\to AA')$ as high as current collider sensitivities would surely require a large amount of tuning.

In light of this, we suggest that it might be worth
reconsidering the necessity of searching for the Higgs
boson decaying to a photon and a dark photon \cite{Beauchesne:Inprep}.

{\it Acknowledgments -- }
This work was supported by the Ministry of Science and Technology of Taiwan under Grants No. MOST-108-2112-M-002-005-MY3 and No. NSTC-111-2112-M-002-018-MY3 and National Center for Theoretical Sciences, Taiwan.

\bibliography{biblio}
\bibliographystyle{utphys}

\end{document}